\documentclass[doublecol]{epl2} 

\title{A test of the bosonic spinon theory for the triangular antiferromagnet spectrum} 

\author{A. Mezio, C. N. Sposetti, L. O. Manuel and A. E. Trumper}
\shortauthor{A. Mezio \etal}

\institute{Instituto de F\'{\i}sica Rosario (CONICET) and Universidad Nacional de Rosario, Boulevard 27 de Febrero 210 bis, 
(2000) Rosario, Argentina}

\abstract{We compute the dynamical structure factor of the spin-$\frac{1}{2}$ triangular Heisenberg model using the 
mean field Schwinger boson  theory. We find that a reconstructed dispersion, resulting from a non trivial redistribution 
of the spectral weight, agrees quite well with the spin excitation spectrum recently found with series expansions. 
In particular, we recover  the strong renormalization with respect to linear spin wave theory along with the appearance of 
roton-like minima. Furthermore, near the roton-like minima the contribution of the two spinon continuum to the static 
structure factor is about $40\%$ of the total weight. By computing the density-density dynamical structure factor, we 
identify an unphysical weak signal of the spin excitation spectrum with the relaxation of the local constraint of the 
Schwinger bosons at the mean field level. 
Based on the accurate description obtained for the static and dynamic ground state properties, we 
argue that the bosonic spinon theory should be considered seriously as a valid alternative to interpret the physics 
of the triangular Heisenberg model.}

\begin{document}

\maketitle

\section{Introduction}

During a long time the magnetic ground state of the spin-$\frac{1}{2}$ triangular Heisenberg model (THM) has attracted the 
attention of many researchers, due to the possible realization of the resonating valence bond (RVB) ground state proposed by 
P. W. Anderson in $1973$ \cite{Anderson73}. The revival of the RVB theory for the cuprates \cite{Anderson87} prompted the 
investigations of quantum disordered ground states within large $N$ theories where the Heisenberg interaction is naturally 
written in terms of singlet bond operators  and fractional spin-$\frac{1}{2}$ excitations with  bosonic or fermionic 
character \cite{Arovas88}. The fermionic version leads to exotic disordered ground states \cite{Affleck88} while the bosonic 
one allows to describe disordered and ordered ground states \cite{Read91} by relating the magnetization with the condensation 
of bosons \cite{Chandra90}. For this case, using gauge field theoretical arguments, it has been conjectured that, when short 
range spiral correlations are present in the disordered phases, the bosonic spinons would be in a deconfined 
regime \cite{Read91}. Therefore, a broad two spinon continuum is expected in the spin excitation spectrum.

From the numerical side, instead, thanks to the enormous effort of the community to develop unbiased  
techniques \cite{Bernu92,Capriotti99,Numerics,Weihong06b}, it has been firmly established that the ground state  of the 
spin-$\frac{1}{2}$ THM is a robust $120^{\circ}$ N\'eel order. These numerical results precluded the fermionic version of 
the RVB theory, giving support to both the linear spin wave theory (LSWT) and the bosonic version of the RVB  theory, 
namely the Schwinger boson theory. In fact, both theories agree quite well with numerical results on finite size 
systems \cite{Manuel98,Lecheminant95}, although for spiral phases the singlet structure of the mean field Schwinger 
bosons theory does not recover the spin wave 
dispersion relation in the large $s$ limit \cite{Chandra91}. Consequently, linear spin wave theory seemed to capture the 
quantum and semiclassical features expected for a $120^{\circ}$ N\'eel ground state  of the THM.
However, recent series expansions studies \cite{Weihong06,Weihong06b} challenged LSWT, showing that 
for $s\!=\!\frac{1}{2}$ the functional form of the dispersion relation differs considerably (points of fig. \ref{fig2}) 
from that of LSWT (solid line of fig. \ref{fig2}). 
In particular, it was observed a strong downward renormalization of the high energy part of the spectrum along with the 
appearance of roton-like minima at the midpoints of edges of the hexagonal Brillouin zone (BZ) (B and D points of the 
inset of fig. \ref{fig1}). The authors argued that the differences with LSWT could be attributed, probably, to the presence 
of fermionic spinon excitations. Nevertheless, further spin wave  studies \cite{Starykh06} showed that, to first order 
in $1/s$, there appear non trivial corrections to the linear spin wave dispersion due to the non collinearity of the ground 
state, giving a fairly accurate description of the series expansion results. However, magnons are not well defined for an 
ample region of the BZ \cite{Chernyshev06}. Another question, regarding the spectrum of the THM, is the nature of the 
multiparticle continuum above the one magnon states. For instance, it is believed that the broad multiparticle continuum 
measured in the  $Cs_2CuCl_4$ compound is better described by an interacting spinon picture than a magnon one \cite{Coldea03}.  
In this sense, given accurate predictions of the Schwinger boson theory for the static ground state properties of the 
THM \cite{Manuel98}, it is important to investigate  whether the anomalous features of the spectrum  found with series 
expansions can be captured, or not, by this alternative theory that naturally incorporates fractional spin-$\frac{1}{2}$ 
excitations.

In the present article we investigate the validity of the bosonic spinon theory to interpret the spin 
excitation spectrum of the spin-$\frac{1}{2}$ THM. Our main finding is that the mean field Schwinger boson theory 
(intensity curves of fig. \ref{fig2}), based on the two singlet operator scheme \cite{Ceccatto93}, reproduces qualitatively 
and quantitatively quite well the recent series expansions results. By computing the dynamical structure factor, we 
remarkably find that the expected spin excitation spectrum is recovered by a reconstruction resulting from a non trivial 
redistribution of the spectral weight located at the spinonic branches shifted by $\pm \frac{\bf Q}{2}$, 
where ${\bf Q}=(\frac{4}{3}\pi,0)$ is the magnetic wave vector. By computing the density-density dynamical structure 
factor, we were able to identify, at the mean field level, the remnant weaker signal of the spectrum with the relaxation of the 
local constraint of the number of bosons. We also discuss the validity of the alternative mean field decoupling based on 
one singlet operator scheme.

\section{Mean field Schwinger bosons approximation}

In the Schwinger boson representation \cite{Arovas88} the spin operators are expressed as 
${\hat{\bf S}}_i\!\!=\! \frac{1}{2}{\bf b}^{\dagger}_i \vec{\sigma} \;{\bf b}_i$, with the spinor
${\bf b}^{\dagger}_i \!=\!(\hat{b}^{\dagger}_{i\uparrow}; \hat{b}^{\dagger}_{i\downarrow})$ composed by the bosonic 
operators $\hat{b}^{\dagger}_{i\uparrow}$ and $\hat{b}^{\dagger}_{i\downarrow}$, and
$\vec{\sigma}\!\!=\!\!(\sigma^x,\sigma^y,\sigma^z)$ the Pauli matrices. To fulfil the spin algebra the constraint 
of $2s$ bosons per site, $\sum_{\sigma}\hat{b}^{\dagger}_{i\sigma}\hat{b}_{i\sigma}\!=2s$, must be imposed. Then, 
the spin-spin interaction of the Heisenberg Hamiltonian can be written as 
\begin{equation}
\hat{{\bf S}}_i \!\cdot \! \hat{{\bf S}}_j=\;: \hat{B}^{\dagger}_{ij} \hat{B}_{ij}:  -\hat{A}^{\dagger}_{ij}\hat{A}_{ij},
\label{int}
\end{equation}
where $::$ means normal order and the singlet bond operators are defined as 
$\hat{A}^{\dagger}_{ij}\!=\!\frac{1}{2}\sum_{\sigma}\sigma \hat{b}^{\dagger}_{i \sigma}
\hat{b}^{\dagger}_{j \bar{\sigma}}$ and 
$\hat{B}^{\dagger}_{ij}\!=\!\frac{1}{2}\sum_{\sigma}\hat{b}^{\dagger}_{i\sigma}\hat{b}_{j \sigma}$. 
We will briefly describe the main steps of the mean field while the details of the calculation can be found 
in our previous works \cite{Manuel98, Ceccatto93}. Introducing a Lagrange multiplier $\lambda $ to impose the 
local constraint on average  and performing a mean field decoupling of 
eq. (\ref{int}), such as $A_{ij}=\langle\hat{A}_{ij}\rangle=\langle\hat{A}^{\dagger}_{ij}\rangle$ 
and $B_{ij}=\langle\hat{B}_{ij}\rangle=\langle\hat{B}^{\dagger}_{ij}\rangle$, the diagonalized mean field Hamiltonian results
$$\hat{H}_{MF}= E_{\texttt{gs}}+\sum_{\bf k} \omega_{\bf k} \left[\hat{ \alpha}^{\dagger}_{{\bf k}\uparrow} 
\hat{\alpha}_{{\bf k}\uparrow}+
\hat{\alpha}^{\dagger}_{-{\bf k}\downarrow} \hat{\alpha}_{-{\bf k}\downarrow} \right],$$
where 
$$E_{\texttt{gs}}=\frac{1}{2}\sum_{\bf k}\omega_{\bf k}+ \lambda N (s+\frac{1}{2})$$
is the ground state energy and  
$$
\omega_{{\bf k}\uparrow}=\omega_{{\bf k}\downarrow}= \omega_{\bf k}=[(\gamma^B_{\bf k}+\lambda)^2- 
(\gamma^A_{\bf k})^2]^{\frac{1}{2}},
$$
is the spinon dispersion relation with geometrical factors, 
$ \gamma^B_{\bf k}\!=\! \frac{1}{2} J\sum_{\delta}  B_{\delta} \cos {\bf k}. \delta$ and 
$\gamma^A_{\bf k}\!=\! \frac{1}{2} J\sum_{\delta} A_{\delta} \sin {\bf k}. \delta$, and 
with the sums going over all the vectors $\delta$ connecting the first neighbours of a triangular lattice. 
The mean field parameters has been chosen real  and satisfy the relations $B_{\delta}\!=\!B_{-\delta}$ and 
$A_{\delta}\!=\!-A_{-\delta}$. The ground state wave function of $\hat{H}_{MF}$ 
can be written in a Jastrow form \cite{Chandra90},
\begin{equation}
|\texttt{gs}\rangle = \exp \left[ \sum_{ij} f_{ij} \hat{A}^{\dagger}_{ij}\right]
|0\rangle_b ,
\label{WFreal}
\end{equation}
where $|0\rangle_b$ represents the vacuum of Schwinger bosons and the odd pairing function 
is defined as  $f_{ij}\!\!=(\frac{1}{N}) \sum_{\bf k} f_{\bf k} e^{\imath {\bf k}({\bf r}_i-{\bf r}_j)}$,  
with $f_{\bf k}\!\!=\!\! -v_{\bf k}/u_{\bf k}$ , and Bogoliubov coefficients $u_{\bf k}\!= \![\frac{1}{2}(1+
\frac{\gamma^B_{\bf k}+\lambda}{\omega_{\bf k}} )]^{\frac{1}{2}}$ and  $v_{\bf k}\!= \!\imath\ {\it sgn}(
\gamma^A_{\bf k})[\frac{1}{2}(-1+\frac{\gamma^B_{\bf k}+\lambda}{\omega_{\bf k}} )]^{\frac{1}{2}}$. The singlet 
bond structure of eq. (\ref{WFreal}) guarantees the singlet behavior of $|\texttt{gs}\rangle$. Even if the 
Lieb-Mattis theorem cannot be applied to non bipartite lattices, the singlet character of the ground state for 
cluster sizes with an even number of sites $N$ has been confirmed numerically \cite{Bernu92,Capriotti99}. It should 
be noted, however, that $|\texttt{gs}\rangle$ is not a true RVB state because the constraint is only satisfied on 
average. Furthermore, by solving the self consistent mean field equations at zero temperature, 
\begin{eqnarray}
A_{\delta}&=& \frac{1}{2N}\sum_{\bf k} \frac{\gamma^A_{\bf k}}{\omega_{\bf k}} \sin{\bf k}. {\delta} \nonumber \\
B_{\delta}&=& \frac{1}{2N}\sum_{\bf k} \frac{(\gamma^B_{\bf k}+\lambda)}{\omega_{\bf k}} \cos{\bf k}. {\delta} \label{self} \\  
s+\frac{1}{2}&=& \frac{1}{2N}\sum_{\bf k} \frac{(\gamma^B_{\bf k}+\lambda)}{\omega_{\bf k}}, \nonumber
\end{eqnarray}
it is found 
that as the system size $N$ increases the singlet ground state $|\texttt{gs}\rangle$ develops $120^{\circ}$ N\'eel 
correlations signalled by the minimum gap of the spinon dispersion located at 
$\pm\frac{\bf Q}{2}$, where ${\bf Q}\!=\!(\frac{4}{3}\pi,0)$ is the magnetic wave vector \cite{Note1}. 
As the spinon gap behaves as $\omega_{\pm\frac{\bf Q}{2}}\!\sim\! 1/N$, for large system sizes the singular 
modes  of eq. (\ref{self}) can be treated apart, analogously to a Bose condensation phenomena \cite{Chandra90}. 
In particular, the local magnetization $m({\bf Q})$ can be derived from the last line of eq. (\ref{self}), 
yielding the relation \cite{Hirsch89} 
$$\frac{1}{2N}\frac{(\gamma^B_{\frac{\bf Q}{2}}+\lambda)^2}{\omega^2_{\frac{\bf Q}{2}}}=S({\bf Q})= \frac {N}{2}m^2({\bf Q}),$$
where $S({\bf k})\!\!=\!\!\!\sum_{\bf R}  e^{\imath {\bf k}. {\bf R}}\langle\texttt{gs}|\hat{S}_0 \!\!\cdot\! \!\hat{S}_{\bf R} 
|\texttt{gs}\rangle$ is the static structure factor. Formally, it can be shown that in the thermodynamic limit 
$|\texttt{gs}\rangle$ is degenerated with a manifold of Bose condensate ground states, each one corresponding to all the 
possible orientations, in spin space, of the $120^{\circ}$ N\'eel order. In the Schwinger boson language the condensate of 
the up/down bosons at $\pm \frac{\bf Q}{2}$  and the normal fluids of bosons corresponds to the spiralling 
magnetization $m({\bf Q})$ and the zero point quantum fluctuations, respectively \cite{Chandra90}. For the triangular lattice 
the present mean field approximation \cite{Gazza93} gives a local magnetization $m=0.275$.

An alternative procedure,  is to use the operator identity 
$:\!\!\hat{B}^{\dagger}_{ij}\hat{B}_{ij}\!\!:\!\!+\hat{A}^{\dagger}_{ij}\hat{A}_{ij}\!\!=\!\!S^2$,  and 
write the spin-spin interaction (\ref{int}) in terms of the singlet operator 
$\hat{A}_{ij}$ \cite{Arovas88,Yoshioka91,Sachdev92}:
\begin{equation}
\hat{\bf S}_i \!\cdot\! \hat{\bf S}_j= -2\hat{A}^{\dagger}_{ijd}\hat{A}_{ij}+S^2.
\label{AA}
\end{equation}
Even if eqs. (\ref{int}) and (\ref{AA}) are equivalent, the latter leads to a different mean field 
decoupling with parameters $A_{\delta}$ and $\lambda$ \cite{Yoshioka91,Sachdev92}. 
In table \ref{table1} it is shown the values of the ground state energy and magnetization for the 
THM obtained  with the two  mean field Schwinger boson decouplings  along with Gaussian fluctuations \cite{Manuel98}, 
linear spin wave theory, non linear spin wave theory (LSWT$+1/s$) \cite{Chernyshev09}; 
and quantum Monte Carlo \cite{Capriotti99} (QMC) results \cite{Note2}. 
Even though it has not yet been calculated,  
we expect that Gaussian fluctuations  above the mean field will reduce the magnetization, as has already been found 
for the spin stiffness in the THM \cite{Manuel98}. From table \ref{table1}  it is seen that the two singlet scheme describe 
quantitatively better the static properties  of the THM. 
\begin{table}
\caption{Energy and magnetization
of the $120^{\circ}$ N\'eel ground state of the spin-$\frac{1}{2}$ Heisenberg
antiferromagnet on the triangular lattice as obtained
with  mean field Schwinger bosons within one \cite{Yoshioka91} ($A$) and two \cite{Gazza93} ($A B$) singlet scheme;
Gaussian fluctuations \cite{Manuel98} above the $AB$ mean field ($AB+\textrm{Fluct}$), 
Quantum Monte Carlo \cite{Capriotti99} (QMC), 
linear spin wave theory (LSWT)  and non linear spin wave theory (LSWT$+1/s$) \cite{Chernyshev09}}.
\label{table1}
\begin{center}
\begin{tabular}{lllll}
\hline \hline 
& & $E/JN $ & & $m$ \\
 \hline
 $A$ & & -0.7119 & & 0.328 \\
 $AB$ & & -0.5697 & &0.275 \\
 $AB+$\textrm{Fluct} & &  -0.5533 & &  \\
QMC  & & -0.5458(1) & & 0.205(1) \\ 
LSWT & & -0.5388 & & 0.2387\\
 LSWT$+1/s$ & & -0.5434 & & 0.2497 \\
\hline \hline
 \end{tabular}
\end{center}
\end{table}

\section{Dynamical structure factor}

\subsection{Spin-spin correlation functions}

We study the spectrum through the dynamical structure factor at $T=0$, 
defined as
$$
S^{\alpha \alpha}\!({\bf k},\omega)= \sum_{n}  |\langle\texttt{gs}|\hat{\bf S}^{\alpha}_{\bf k}(0)|n\rangle|^2 \delta 
(\omega-(\epsilon_n-E_{\texttt{gs}}))\nonumber, 
$$
where $\alpha$ denotes $x,y,z$, $|n\rangle$ are the excited states, and 
$\hat{\bf S}^{\alpha}_{\bf k}$ is the Fourier transform of $\hat{\bf S}^{\alpha}_i$. As we work on finite systems 
the $SU(2)$ symmetry is not broken explicitly and $S^{x x}\!\!=\!\!S^{y y}\!\!=\!\!S^{z z}$ (in what follows 
the $\alpha\alpha$ indices are discarded). A straightforward calculation leads to the expression 
\begin{equation}
S\!({\bf k},\omega)\!=\!\frac{1}{4N}\!\!\sum_{{\bf q}} |u_{{\bf k}+{\bf q}} v_{\bf q} - u_{{\bf q}} v_{{\bf k}+{\bf q}}|^2 
\delta (\omega-(\omega_{-{\bf q}}+\omega_{{\bf k}+{\bf q}})),
\label{Skw}
\end{equation}
which satisfies the correct sum rule $\int\! \sum_{{\bf k}\alpha}S^{\alpha \alpha}({\bf k},\omega)d\omega= Ns(s+1)$. 

As at the mean field level the triplet excitations are made of two spin-$\frac{1}{2}$ free spinons a broad 
two spinon continuum is expected. Nevertheless, as the $120^{\circ}$ long range N\'eel order is developed there can be 
distinguished three distinct contributions in the spectrum. Following the interpretation of the 
spectra of \cite{Lefmann94}, it is instructive to split eq. 
(\ref{Skw}) as 
$$S({\bf k},\omega)= S^{sing}_{{\bf k},\omega}+ S^{cont}_{{\bf k},\omega},$$
by using the fact that $u_{\pm \frac{\bf Q}{2}}\!=\!|v_{\pm \frac{\bf Q}{2}}|\!\sim\!(\frac{Nm}{2})^{\frac{1}{2}}$ and 
$\omega_{\pm{\frac{\bf Q}{2}}}\sim0$. For ${\bf k}=\pm{\bf Q}$, the spectrum is dominated by zero energy processes that 
create two spinons in the condensate. This gives rise to the magnetic Bragg peaks which, to leading order, behave as 
$ S^{sing}_{\pm{\bf Q},\omega}\!\!\sim Nm^2\delta(\omega)$. For ${\bf k}\neq \pm{\bf Q}$, the spectrum is dominated by 
low energy processes that create one spinon  in the condensate and another one in the normal fluid. This gives rise to a 
double peaked signal proportional to $m$, represented by 
\begin{eqnarray}
 S^{sing}_{{\bf k},\omega}\!&=&\! \frac{m}{4} |\imath \;u_{{\bf k}+\frac{\bf Q}{2}}\! -\! v_{{\bf k}+\frac{\bf Q}{2}}|^2 
\delta (\omega-\omega_{{\bf k}\!+\!\frac{\bf Q}{2} }\!)+ \nonumber \\ 
&+&\frac{m}{4} |\imath\; u_{{\bf k}\!-\!\frac{\bf Q}{2}}\! +\! v_{{\bf k}-\frac{\bf Q}{2}}|^2 \delta (\omega-
\omega_{{\bf k}-\frac{\bf Q}{2}}\!).\nonumber
\end{eqnarray}
Then, the shifted spinon dispersion $\omega_{{\bf k}\pm\frac{\bf Q}{2}}$ can be identified with  the low energy physical 
magnetic excitations. Finally, at high energy, the spectrum is dominated by the processes of creating two spinons in the 
normal fluid. This gives rise to a broad continuum represented by  
$$
S^{cont}_{{\bf k},\omega}\!=\!\!\frac{1}{4N}\!\!\sum_{\bf q} {}^{'} \!|u_{{\bf k}+
{\bf q}} v_{\bf q} - u_{{\bf q}} v_{{\bf k}+{\bf q}}|^2 \delta (\omega-(\omega_{-{\bf q}}+
\omega_{{\bf k}+{\bf q}})),\nonumber
$$
where the prime means that sum goes over the triangular BZ except for ${\bf q}= \pm \frac{\bf Q}{2} $ or 
$\pm \frac{\bf Q}{2}-{\bf k}$.

\begin{figure}[h]
\onefigure[width=0.3\textwidth,angle=-90]{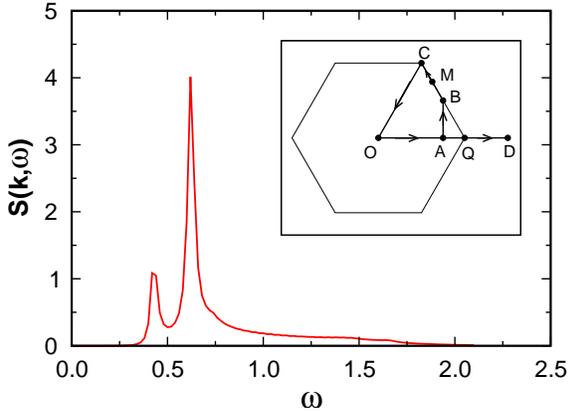}
\caption{Dynamical structure factor, $S({\bf k},\omega)$, for momentum $M=(\frac{5}{6}\pi,\frac{\sqrt{3}}{2}\pi)$. 
Inset: path of the triangular BZ along which the spectrum has been investigated. 
$O\!\!=\!\!(0,0)$, $A\!\!=\!\!(\pi,0)$, $Q\!=\!(\frac{4}{3}\pi,0)$, $D\!=\!(2\pi,0), B\!=\!(\pi,\frac{1}{\sqrt{3}}\pi)$, 
and $C\!=\!(\frac{2}{3}\pi,\frac{2}{\sqrt{3}}\pi)$. $\omega$ is measured in units of $J$.}
\label{fig1}
\end{figure}

\begin{figure}[t]
\onefigure[width=0.29\textwidth,angle=-90]{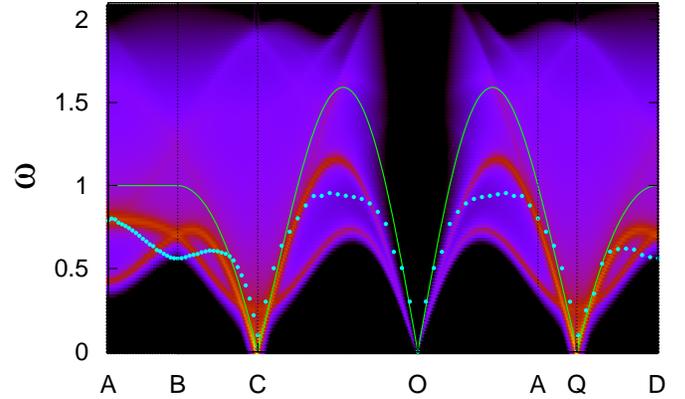}
\caption{Intensity curves for the dynamical structure factor, $S({\bf k},\omega)$, calculated  with 
the mean field Schwinger bosons theory within the two singlet scheme. Solid green line and blue points 
are the dispersion relations obtained with LSWT and series expansions \cite{Weihong06}, respectively. 
The path along the BZ is shown in the inset of fig. \ref{fig1}.}
\label{fig2}
\end{figure}

In fig. \ref{fig1} we have plotted eq. (\ref{Skw}) for the  $M$ point of the BZ 
(see inset of fig. \ref{fig1}). As noticed above, the low energy double peaked structure comes from 
$S^{sing}_{{\bf k},\omega}$ while the high energy tail corresponds to the continuum 
$S^{cont}_{{\bf k},\omega}$. In order to get the spectrum in the energy-momentum space
we have plotted in fig. \ref{fig2} the intensity curves of  $S({\bf k},\omega)$ (eq. (\ref{Skw})) 
along the path shown in the inset of fig. \ref{fig1}. The yellow and red curves are the shifted 
spinon dispersion $\omega_{{\bf k}\mp\frac{\bf Q}{2}}$ of $S^{sing}_{{\bf k},\omega}$  while 
the blue zone corresponds to $S^{cont}_{{\bf k},\omega}$. In the figure we compare with the  
dispersion relations obtained with LSWT (solid line) and  the recent series expansion 
calculations \cite{Weihong06} (points). At low energies the dispersion agrees quite well 
with LSWT and series expansions, being the spectral weight mostly located around 
${\bf k}\sim \pm{\bf Q}$ (points Q and C). In this regime the physical excitations correspond 
to long range transverse distorsions of the local magnetization which are correctly described 
by both, LSWT and mean field Schwinger bosons. At higher energies LSWT is not valid any more 
since the true spin excitations show a strong downward renormalization along with the appearance 
of roton-like minima (points). Remarkably, the  mean field Schwinger boson theory predicts 
a non trivial redistribution of the spectral weight between the two spinon branches modulated 
by the form factor of eq. (\ref{Skw}). The reconstructed dispersion, resulting from those 
pieces of spinon dispersion with the dominant spectral weight, reproduces quite well the 
series expansions results. In particular, the crossing of the spinon dispersions at points 
$B$ and $D$  can be identified with the roton-like minima observed in series expansions. 
Regarding the interpretation of the roton minima, the singlet bond structure of the Schwinger 
boson theory takes naturally into account the collinear spin fluctuations even in the presence of the $120^{\circ}$ 
N\'eel order of the THM. For instance, the roton minimum located at $B$ can be interpreted  as the 
development of magnetic correlations  modulated by the magnetic wave vector $(\pi,\frac{1}{\sqrt{3}}\pi)$  
which corresponds to certain  collinear correlations pattern, while the other two non equivalent midpoints 
of the edges of the hexagonal BZ corresponds to different  collinear fluctuations patterns. In fact, if 
these fluctuations are favoured by introducing spatially anisotropic or second neighbours exchange 
interactions  the roton minima soften, giving rise to the new Goldstone mode structure of the stabilized 
collinear ground state \cite{Gazza93,Manuel99}.

\begin{figure}[ht]
\onefigure[width=0.3\textwidth,angle=-90]{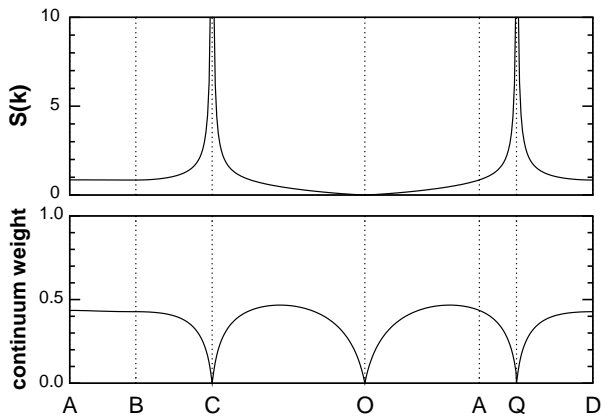}
\caption{Static structure factor (upper panel) and relative weight of the two spinon continuum, 
$\int  S^{cont}_{{\bf k},\omega}/ S({\bf k})d\omega$, (bottom panel) along the same path of the BZ.}
\label{fig3}
\end{figure}

Performing the frequency integration it is  possible to analyze the relative weight of the two 
spinon continuum (blue zone of fig. \ref{fig2}) to the static structure factor $S({\bf k})$ \cite{Capriotti05}.
In fig. \ref{fig3} we plot $S({\bf k})$ with diverging peaks located at the expected magnetic wave vectors $\pm{\bf Q}$ 
(upper panel), along with the  relative weight of the two spinon continuum, 
$\int S^{cont}_{{\bf k},\omega}/ S({\bf k}) d\omega$ (bottom panel). Interestingly, the contribution of the 
two spinon continuum to $S({\bf k})$ is neglegible around $\pm{\bf Q}$ while outside their neighbourhood, and 
in particular at the roton position, the  contribution to $S({\bf k})$  is about $40\%$.

\subsection{Density-density correlation functions}

The small peak of fig. \ref{fig1} leads to the remnant weak signal of fig. \ref{fig2} which can be traced 
back to the local density fluctuation of Schwinger bosons. In fact, to describe the physical Hilbert space 
of the spin operators the local constraint of the Schwinger bosons must be satisfied exactly, 
$\hat{{\bf S}}^2_i=\frac{n_i}{2}(\frac{n_i}{2}+1)$. Then, no fluctuations on the number of boson per site 
should be observed.  However, since the constraint is taken into account on average there are unphysical spin 
fluctuations in $S({\bf k},\omega)$ coming from such density fluctuations. In order to identify them we have 
computed the density-density  dynamical structure factor defined as
$$
\emph{N}({\bf k},\omega)= \sum_{n} \! |\langle\texttt{gs}|\hat{n}_{\bf k}(0)|n\rangle|^2 \delta 
(\omega-(\epsilon_n-E_{\texttt{gs}}))\nonumber,
$$
where $\hat{n}_{\bf k}$ is the Fourier transform of the number of bosons per site, 
$\hat{n}_i=\sum_{\sigma}\hat{b}^{\dagger}_{i\sigma}\hat{b}_{i\sigma}$. A little of algebra leads to the expression
\begin{equation}
\emph{N}({\bf k},\omega)\!=\frac{1}{N}\!\!\sum_{{\bf q}} |u_{{\bf k}+
{\bf q}} v_{\bf q} +u_{{\bf q}} v_{{\bf k}+{\bf q}}|^2
\delta (\omega-(\omega_{-{\bf q}}+\omega_{{\bf k}+{\bf q}})),
\label{Nkw}
\end{equation}

\begin{figure}[t]
\onefigure[width=0.29\textwidth,angle=-90]{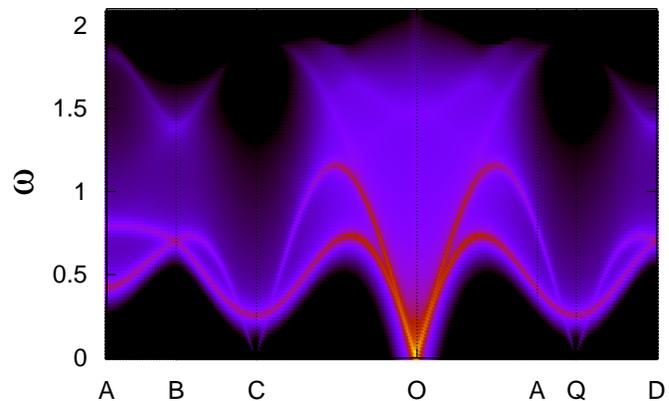}
\caption{Intensity curves for the density-density dynamical structure factor, $\emph{N}({\bf k},\omega)$, calculated 
within the mean field Schwinger bosons based on the two singlet scheme. The path along the BZ is shown in the inset 
of fig. \ref{fig1}.}
\label{fig4}
\end{figure}
which is similar to  eq. (\ref{Skw}), except to the plus sign within the form factor. 
If we split the two spinon contributions as  $\emph{N}({\bf k},\omega)= \emph{N}^{sing}_{{\bf k},\omega}+ 
\emph{N}^{cont}_{{\bf k},\omega}$ it is easy to show that the main signal is located again at the shifted 
spinon dispersions $\omega_{{\bf k}\mp\frac{\bf Q}{2}}$. But now, due to the different form factor, there is an 
important spectral weight transfer between such spinon dispersions. This is shown in fig. \ref{fig4} where we have 
plotted the intensity curves of $\emph{N}({\bf k},\omega)$ (eq. (\ref{Nkw})). It can be clearly observed  that now the 
dominant  signal is gapped at Q and C points,  while most of the spectral weight is located around ${\bf k}\sim0$. Such 
a soft mode can be identified with a spurious tendency of the bosonic system to phase separation. Given the notable 
resemblance  with the strong signal of $\emph{N}({\bf k},\omega)$, we suggest that the low energy weak signal 
of figs. \ref{fig1} and \ref{fig2} could be ascribed with the unphysical density fluctuation effects which we expect 
to disappear once they are projected out. For the unfrustrated square lattice 
$\omega_{{\bf k}+(\frac{\pi}{2},\frac{\pi}{2})}=\omega_{{\bf k}-(\frac{\pi}{2},\frac{\pi}{2})}$ 
so both, the unphysical and the physical spin excitations, overlap in energy-momentum space, giving rise only to one low 
energy band in $S({\bf k},\omega)$ \cite{Arovas88,Capriotti05}. 

\subsection{Comparison with the one singlet scheme}

\begin{figure}
\onefigure[width=0.3\textwidth,angle=-90]{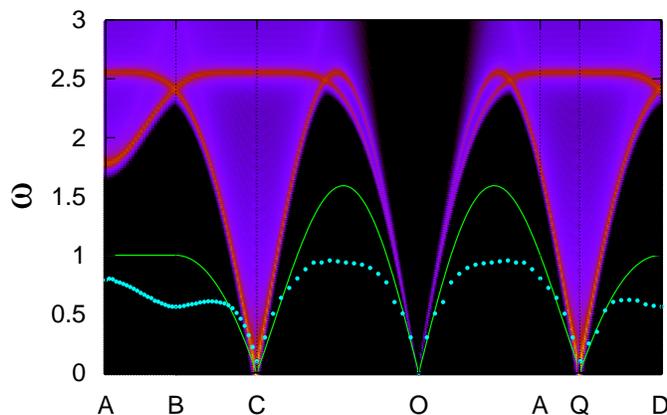}
\caption{Intensity curves for the dynamical structure factor calculated within the mean field Schwinger bosons 
based on the one singlet scheme. The path along the BZ is shown in the inset of fig. \ref{fig1}. Solid line 
and points are the same as in fig. \ref{fig2}.}
\label{fig5}
\end{figure}

So far we have found that the mean field Schwinger boson within the two singlet scheme reproduces quite well the 
series expansions spectrum. It is also interesting to compare with the predictions of the one singlet scheme, 
since it is widely used in the literature.
The first difference is the incorrect sum rule 
$\int \! \sum_{{\bf k}\alpha}S^{\alpha\alpha}({\bf k},\omega)d\omega= \frac{3}{2}Ns(s+1)$ which implies the well 
known $\frac{2}{3}$ factor of Arovas and Auerbach \cite{Arovas88}. Furthermore, in fig. \ref{fig5},  we have 
computed $S({\bf k},\omega)$ after solving the corresponding self consistent equations for 
the parameters $A_{\delta}$, and $\lambda$. At very low energies the spectrum seems to be correct around 
points $C$, $O$ and $Q$. However, at higher energies it is impossible to discern 
a reconstructed dispersion that fit the series expansion results along the whole path of the BZ, besides the 
factor about $3$ in the energy scale. Therefore, we conclude that the two singlet scheme turns out the proper 
framework to describe correctly 
the spectrum of the THM.
Besides its quantitative accuracy, there are symmetry arguments that give further support to the two singlet 
scheme. In the literature, the one singlet scheme has been justified  as the saddle point of a symplectic 
$Sp(N)$ theory,  originally adapted to extend previous  large $N$ works \cite{Arovas88} to non bipartite 
lattices \cite{Read91}. More recently, however, Flint and Coleman \cite{Flint08}  demonstrated that if the 
$\hat{B}_{ij}$  and $\hat{A}_{ij}$ operators are kept the corresponding large $N$ extension preserves the time 
reversal properties of the spins, in contrast to the $Sp(N)$ theory. Finally, it is worth to stress that the two 
singlet scheme is the basis of the $Z_2$ spin liquid theory, specially formulated to describe magnetically disordered 
phases \cite{Wang06}.

\section{Conclusions}

We have demonstrated that the singlet structure of the mean field ground state along with the fractional character 
of the spin excitations of the Schwinger boson theory take naturally into account the anomalous excitations of the 
spin-$\frac{1}{2}$ triangular Heisenberg model recently observed \cite{Weihong06,Weihong06b}. The appearance of the 
roton-like minima can be attributed to the tendency of the magnetic ground state to be correlated collinearly, even 
in the presence of $120^{\circ}$ N\'eel order. By computing the density-density dynamical structure factor, and 
thanks to the series expansion results,  we were 
able for the first time to discern, at the mean field level, 
between the physical and the spurious fluctuations coming from the relaxation of 
the local constraint. A further investigation within the context of the Schwinger boson theory reveals that the 
correct description of the spectrum depends crucially on the mean field decoupling. In particular, the two singlet 
scheme turns out more appropriate than the one singlet scheme. 
Based on the accurate description of the ground state static properties \cite{Manuel98} (see table \ref{table1})
and in the light of the present results for the spectrum, we think that the bosonic spinon hypothesis should be 
considered seriously as an alternative viewpoint to interpret the physics of the triangular Heisenberg model. 
At the mean field level the triplet excitations consist of two spin-$\frac{1}{2}$ free spinons and, 
besides the low energy bands due to the onset of the long range order, there is a broad two spinon 
continuum, which could be related with the magnon decay found in the literature \cite{Chernyshev09}. In this sense,
it would be important to improve the present mean field theory by deriving an effective interaction between 
spinons resulting from $1/N$ corrections or a better implementation of the constraint. We would expect a picture 
of tightly bound spinons near the Goldstone modes while at high energies they would be weakly bound. 
Work in this 
direction is in progress. Finally, we hope our present analysis in terms of bosonic spinons could help for a better 
understanding of the unconventional neutron scattering spectra of the $Cs_2CuCl_4$ compound \cite{Coldea03}. 

\acknowledgments
We thank W. Zheng and R. Coldea for sending us their series expansions results, and C. Lhuillier and C. Batista for 
very useful discussions. This work was supported by PIP2009 under grant  No. $1948$.

\end{document}